\documentclass[12pt]{article}
\usepackage{amsmath}
\usepackage{amsfonts}
\usepackage{graphicx}
\usepackage{amssymb}
\usepackage{yhmath}
\usepackage{pdfpages}
\graphicspath{ {images/} }

\newtheorem{Def}{Definition}[section]

\newcommand{\re}{\mathbb{R}}

\def \seq {\subseteq}

\def \qed {\hfill \vrule height6pt width 6pt depth 0pt}
\def \noi {\noindent}

\begin{document}

\title{On accessibility of the core in mutidimensional spatial majority voting situations}
\vspace{0.1cm}

\author{Anindya Bhattacharya\thanks {Department of Economics and Related Studies, University of York, York, YO10 5DD, United Kingdom; Email: anindya.bhattacharya@york.ac.uk;}~~~Francesco Ciardiello\thanks{Department of Economics and Statistics (DISES), University of Salerno, Fisciano 84084,  Italy; Email: fciardiello@unisa.it.}}

\date{\today}

\maketitle

\maketitle

\begin{abstract}

In this paper we consider situations of (multidimensional) spatial majority voting. We analyze such situations having an even number (greater than or equal to 4) of voters and assume that each voter's preference over the set of policies is ``Euclidean": i.e., each voter has a most preferred ``ideal" policy and the voter's pay-offs from policies decrease as the (Euclidean) distance of the policies from the ideal policy goes up. We confine attention to voting situations for which the core has a single element belonging to the interior of the policy set. It is known that under conditions usual in this literature, this element of the core, generally, is not a Condorcet winner: i.e., there is at least one policy outside the core such that the core-policy cannot be reached from it by just one round of majority domination. We demonstrate that, however, under such conditions, starting from any policy in place, by a finite sequence of majority-dominations (we assume, implicitly, sincere voting) the policy in the core is reached eventually.  

\end{abstract}

\vskip1em

\noi {\em  Keywords:} Spatial Voting Situations; Core; Accessibility.\\

\noi

\newpage

\section{Introduction}

\noi It is quite well-known that for a voting situation if the number of voters changes from being odd to even then, {\em everything else remaining the same}, the properties associated with the voting situation may change. Perhaps the simplest and the most well-known example of such ``may change" occurrences is the following. Take a finite set of voters and a finite set of candidates and suppose that each voter has a strict preference ordering over the set of candidates. Then, if there is an odd number of voters then, with majority rule voting, for any two candidates $x$ and $y,$ either $x$ is socially preferred to $y$ or the converse is true: i.e., the resulting social ordering is decisive between any two candidates. But that may not be the case if there is an even number of voters.

\noi In this paper we explore one such issue, but when the situation is such that the voters vote on a set of multi-dimensional policies (the next Section provides the exact formal details), rather than over a finite set of candidates or policies. In this paper we assume sincere voting by the voters (a literature on ``sophisticated" voting and issues like strategyproofness is there, especially following Le Breton and Sen, 1999).

\noi For several soco-economic environments, the core, as the subset of social states which are immune to profitable deviations by coalitions of agents/voters has been studied quite a lot as predictors of what social states would remain stable. For our environment of multi-dimensional spatial voting, Chapter 5 of Austen-Smith and Banks (1999) contains a summary of some central results analyzing the core of such situations.

\noi In the environment of multi-dimensional spatial voting, when there is an odd number of voters, the core has the following attractive property: if the core is non-empty then, under quite general conditions, contains a single element which is also the Condorcet winner of the voting situation (a precise statement of this property is given in Section 3 below). That is, if a policy outside the core is in place then some majority of voters can profitably move to replace that by the policy in the core. 

\noi But when there is an even number of voters, this feature disappears. Under some regularity assumptions usual in this literature Bhattacharya and Ciardiello (2024) show that with an even number of voters if the core of the voting situation is singleton (and if the core element is in the interior of the policy space) then there is at least one policy outside the core for which no majority of voters can profitably move to replace that by the policy in the core.   (i.e., in other words, a Condorcet winner for this situation does not exist).  

\noi Given the feature above, the following question arises. Suppose there is an even number of voters and each voter's preference is ``Euclidean" (precisely defined in the next Section), perhaps the most popular assumption about the voters' preferences in the related literature. Suppose, again, that the core of the voting situation is singleton and that the core-element is in the interior of the policy space. Consider any policy outside the core. Then, do there exist {\em finitely many, but may be more than one, of profitable moves by different majority coalitions of voters} which eventually lead to the element in the core? In this paper we demonstrate that the answer to this question is affirmative.

\noi A somewhat similar question has been looked into in the respective literature. When the core is empty for such a voting situation then McKelvey (1976) showed that ``it is theoretically possible to design voting procedures, which, starting from any given point, will end up at any other point in the space of alternatives, even at Pareto dominated ones" (a summary of some further development along this line of exploration can be obtained in Chapter 6 of Austen-Smith and Banks, 1999). Our result makes a similar but contrasting point: that when such a voting situation admits a non-empty core then it is theoretically possible to design voting procedures, which, starting from any given point, will end up at the stable point in the core.

\noi Similar questions have been explored for other environments too. As an example, Sengupta and Sengupta (1996; a body of literature follwed this paper) demonstrated that for any transferable utility game in coalitional form with nonempty core, given any allocation outside the core, there is an allocation in the core that is reached by a sequence of coalitional dominations.

\noi The plan of this paper is as follows. The next Section provides the preliminary definitions, notation and the initial ideas. The central result of this paper and its proof are given in Section 3. Section 4 provides a couple of concluding remarks.

\section {Preliminary definitions, notation and the initial ideas}

Let $Z\subseteq {\re}^{k}$ be a convex subset of some finite ($k$-)dimensional Euclidean space with $k > 1$ (in what follows, the underlying topological space is taken to be the entire ${\re}^k$). This set, $Z,$ is identified, for our exercise, to be the feasible set of multi-dimensional policies on which a voter votes. In what follows, because we have to use geometrical arguments quite a bit, we shall often call a policy simply a ``point".\\

\noi Let $N$ be the finite set of players or voters. For a set $A,$ by $\#A$ we denote its cardinality. For any two points $x,y \in {\re}^k,$ by $d(x,y)$ we denote the (Euclidean) distance between these two points.\\

\noi For each $i \in N$ the preferences of $i$ on $Z$ is ``Euclidean" with a distinct ideal point: i.e., for each $i \in N$ there exists a {\em distinct} ${\bar x}_i \in Z,$ the ideal point of $i$ in $Z$, such that for any policy $x \in Z,$ $u_i(x)=-(d (x, {\bar x}_i))^2.$ For some of the discussion below, for stating some things succinctly, we consider this functional form on the entire ${\re}^k:$ i.e., for any point $x \in {\re}^k,$ $u_i(x)=-(d (x, {\bar x}_i))^2.$ In what follows, we assume that no three such ideal points are collinear.\\

\noi The spatial voting situation we consider below is obtained by introducing the method of majority rule voting.
\begin{Def}$[$Domination by Majority Rule$]$
Given $x,y \in {\re^k},$ if there exists a coalition $S \seq N,$ such that $\#S>|N|/2$ and $u_i(x)>u_i(y)$ for each $i \in S$ then we denote this relation as $x \succ_S y.$ If there exists a majority coalition $S$ for which $x \succ_S y$ then we denote that as relation as $x \succ y.$\\
If for any such pair of points $x,y,$ both $x$ and $y$ belong to $Z$: i.e., both are considered to be feasible policies on which the voters vote then, if $x \succ y$ then we say that the policy $x$ dominates the policy $y$ by some majority coalition.
\end{Def}
The collection $G=\langle Z,N,(u_i)_{i \in N} \rangle$ is a {\em spatial voting situation with majority rule} (which we shall often refer below simply as a {\em voting situation} with no possibility of confusion).\\

\noi For any $S \seq N$ and any $x \in Z,$ by $C_S(x)$ we denote the set $\{y \in Z \mid u_i(y) \geq u_i(x)$ for each $i \in S\}.$ For any set $A \seq Z,$ by $cl(A),$ $int(A)$ and $bd(A)$ we denote the closure of $A,$ the interior of $A$ and the boundary of $A$ respectively.\\

\noi Next we get on to the central object of our interest.

\begin{Def}[The core of a voting situation]
The core of a voting situation $G$ is the subset
$
K=\{y\in Z \,:\,\nexists x \in Z \,\,\mbox{such that}\,\, x \succ y \}.
$
\end{Def}

\noi Throughout this paper we would work with voting situations for which $K$ contains a single point $x_0$ belonging to the interior of $Z.$ Further, we assume that for each $i \in N,$ $x_0 \neq {\bar x_i}.$ Given these assumptions within our set-up, if $\#N$ is even then by, e.g., Duggan (2018; p.2) there is a set $\mu \subset (N \times N)$ of $\#N/2$ distinct pairs of voters such that for every pair $(i,j) \in \mu$ the normalized gradient of $i$'s pay-off function at $x_0$ is the negative of the normalized gradient of $j$'s pay-off function at $x_0.$ For any two distinct voters $m,p$ denote by $K_{mp}$ the ``$mp$-core", defined as follows:
$K_{mp}=\{x \in Z \, : \,$ there does not exist $y \in Z$ such that $u_m(y)>u_m(x)$ as well as $u_p(y)>u_p(x)\}.$\\
It is easy to see that for any pair of distinct voters $(m,p),$ $K_{mp}$ is the straight line segment connecting ${\bar x}_m$ and ${\bar x}_p.$ Further, it is also easy to see that given our assumptions above, the single core point $x_0$ lies at the intersection of the straight line segments $\overline{mp}$'s where each pair of voters $(m,p)$ is in $\mu$ (and recall that for every voter $m,$ its pairing voter $p$ (for which $(m,p) \in \mu$) is unique). In what follows, we use the notational convention that, for each such pair $(m,p) \in \mu,$ if $m=i \in N,$ then $p,$ its pairing ideal point is denoted by $i'.$\\
For each $(i,i') \in \mu,$ by $L_i$ denote the straight line in ${\re}^k,$ such that the line segment $\overline{ii'}$ is a subset of $L_i:$ i.e., $L_i$ is the straight line extension into ${\re}^k$ of the line-segment ${\overline{ii'}} \subset Z.$\\

\noi Given $z \in Z,$ we call an $M \subset Z$ {\em star-shaped} with respect to $z \in Z$ if the following hold:\\
$(i)$ There exists a positive integer $l>1$ such that there is a family of subsets $\{C_i\}_{i \in \{1, \ldots,l\}} \subset Z$ for which $M=\cup_{i=1,\cdots,l}C_i;$ and additionally,\\
$(ii)$ for any distinct pair $i,j \in \{1,\ldots,l\},$ $C_i \cap C_j=\{z\}.$\\
  
\noi Next recall that for any straight line $L \subset \re^k$ and any point $y \in \re^k \setminus L,$ the {\em reflection} of $y$ with respect to the line $L$ is the point $y'$ such that $L$ is the perpendicular bisector of the line-segment $\overline{yy'}.$\\

\noi The property we explore in this paper is defined as follows.

\begin{Def}[Accessibility]
For any $x,y \in Z,$ $y \neq x,$ the point $x$ is said to be accessible from point $y$ if there exists $\{w_1, \ldots, w_p\} \subset Z$ such that $w_1=x,$ $w_p=y$ and for each $i \in \{1, \ldots, p\},$ $w_i \succ w_{i+1}.$
\end{Def}

\section {The result and its proof}

If $N,$ the set of voters, contains an odd number of elements then, for voting situations under assumptions less restrictive than those in our set-up, the following result is well-known.\\

\noi {\bf{Proposition $0$}} {\em{Consider a voting situation $G$ for which $\#N$ is an odd positive integer. Suppose further that for $G$ the core $K$ is non-empty. Then there is a unique element in the core, $x_0,$ such that for each $y \in Z,$ $y \neq x_0,$ $x_0 \succ y.$}}\\

\noi A proof of Proposition $0$ is given in Cox (1987, p. 411).\\

\noi Bhattacharya and Ciardiello (2024) obtained a sharply contrasting result as follows, with pay-off functions more general than we use here, when $N,$ the set of voters, contains an even number of elements.\\

\noi {\bf{Proposition $0^\prime$}} {\em{Consider a voting situation $G$ for which $\#N$ is an even positive integer. Suppose further that for $G,$ $Z$ is compact, the core $K=\{x_0\}$ is singleton and the point $x_0$ is in the interior of $Z.$ Assume additionally that for at most one $i \in N$ is it the case that $x_0=\bar{x}_i.$ Then there exists $y \in Z,$ $y \neq x_0,$ for which $x_0$ does not dominate $y.$}}\\

\noi In this paper we show the following.\\

\noi {\bf Proposition} {\em Consider a voting situation $G$ for which $\#N$ is an even positive integer $2n;$ $(n>1).$ Then $x_0$ is accessible from any other $y \in Z.$}\\

\noi Some of the ideas for proving this Proposition are similar to those in McKelvey (1976). and Miller et al. (1989).\\

\noi By $M(x_0)$ we denote the subset of points undominated by the single point $x_0$ in the core: i.e., $M(x_0)=\cup_{S \subset N \mid \#S=n}C_S(x_0).$\\

\noi We start with a Fact and a couple of initial Lemmata.\\

\noi {\bf Lemma 1} {\em The set $M(x_0)$  is star-shaped with respect to $x_0.$.}\\

\noi {\bf Proof.} Note that for any $n$-voter coalition $S,$ $x_0 \in C_S(x_0).$ We show that for any two distinct $n$-voter coalitions $S$ and $T$, $S \neq T,$ $(C_S(x_0) \cap C_T(x_0)) \setminus \{x_0\} =\emptyset.$ Suppose not. Then there exists $x \in Z$ such that $x \neq x_0$ and for each $i \in (S \cup T),$ $u_i(x) \geq u_i(x_0).$ Therefore, since $Z$ is convex, there exists some $y$ in the line segment ${\overline{xx_0}} \subset Z$ such that for each $i \in (S \cup T),$ $u_i(y)>u_i(x_0).$ Since the cardinality of $(S \cup T)$ is at least $(n+1),$ then $y$ majority-dominates $x_0$ which leads to a contradiction. \qed\\  

\noi {\bf Remark 1} Note that by Lemma 1, for any two $n$-voter coalitions $S,T;$ $S \neq T,$ for any point $x \in C_S(x_0)$ and $y \in C_T(x_0),$ $x,y \neq x_0,$ for the straight line segment ${\overline{xy}} \subset Z,$ ${\overline{xy}} \cap (Z \setminus M(x_0)) \neq \emptyset.$\\

\noi {\bf Fact 1} {\em For any two points $y$ and $z$ in $Z,$ denote by $H_{yz}=\{x \in \re^k \mid (y-x)^2=(z-x)^2\}.$ Then, $\#\{i \in N \mid u_i(y) \geq u_i(z)\} \geq n$ if and only if the number of ideal points lying in the closed half-space of $H_{yz}$ that contains $y$ is (weakly) greater than the number of ideal points lying in the closed half-space of $H_{yz}$ that contains $z.$ And therefore, if the number of ideal points lying in the open half-space of $H_{yz}$ that contains $y$ is greater than $n$ then $y \succ z.$}\\

\noi {\bf Proof.} Please see Davis et al. (1972), p. 150. \qed\\

\noi {\bf Lemma 2} {\em Take any pair $(i, i') \in \mu.$ Pick $y \in Z.$ Let $y' \in {\re}^k$ be the reflection of $y$ by the straight line $L_i.$ Pick any $x$ belonging to the open line segment ${\overline{y'y}} \setminus \{y',y\}.$ Then $x \succ y.$}\\

\noi {\bf Proof.} Denote by $y_p$ the point at which the line-segment ${\overline{y'y}}$ is bisected perpendicularly by $L_i.$ Denote by $H$ the hyperplane such that $L_i \subset H$ and for every straight line $L \subset H,$ passing through $y_p,$ the projection of $y$ on $L_i,$ the angle between ${\overline{yy^p}}$ and $L$ is $\pi/2$: i.e., $H$ is the hyperplane perpendicular to $\overline{yy'},$ bisecting $\overline{yy'}.$ Denote by $H(y)$ the open half-space with respect to $H$ that contains $y$ and by $H(y')$ the open half-space that contains $y'.$ Pick any $x$ in the open line segment ${\overline{y'y}} \setminus \{y',y\}.$\\
Consider the following Cases.\\
Case 1: Suppose each of the 2-player core-line-segments the intersection of all of which is the interior core point $x_0$ is a subset of $H.$ Then consider the hyperplane $H_{yx}=\{w \in \re^k \mid (y-w)^2=(w-x)^2\}.$ Then each of the $2n$ ideal points belongs to the open half-space of $H_{yx}$ that contains $x.$ Then, by Fact 1, $x \succ y.$\\
Case 2: Let $(j,j') \in \mu$ be such that $\bar x_j \notin H.$ Since $x_0 \in H,$ $\bar x_{j'} \notin H.$ Therefore, if there are $m<n$ ideal points belonging to $H(y)$ then there are exactly $m$ ideal points belonging to $H(y').$ Note that this $m$ can at most be $n-1.$ Recall that $\bar x_{i'}$ as well as $\bar x_{i}$ belong to $H.$ Then consider the hyperplane $H_{yx}=\{w \in \re^k \mid (y-w)^2=(w-x)^2\}.$ Then at least $n+1$ ideal points belong to the open half-space of $H_{yx}$ that contains $x.$ Then, by Fact 1, $x \succ y.$ \qed\\

\noi {\bf The remainder of the proof of the Proposition}\\

\noi Pick any $z \in M(x_0).$ We provide the remainder of the proof of this Proposition along the following Steps.\\

\noi {\em Step 1:} Denote by $C \seq Z$ the convex hull of the ideal points of the voters: i.e., the convex hull of $\{\bar x_i \mid i \in N\}.$ Denote by $E(C)$ the extreme points of $C.$\\ 
In this Step of the proof we demonstrate the following. For at least one pair $(i,i') \in mu,$ there exists $y \in (C \cap L_i)$ such that either $y$ is accessible from $z$ or $y=z.$\\
Consider the following Cases.\\
Case 1: Suppose $z$ is one of the extreme points of $C.$ Then, by construction of $C,$ $z$ must be $\bar x_i$ for some $i \in N.$ If this $z=\bar x_i$ is one of the extreme points of $C,$ then, since the line-segment ${\overline{{z}{x_0}}} \subset int(C),$ there must exist some $y \in (int(C) \cap L_i)$ (may be infinitesimally close to $z$) with the following property:\\
there exists a hyperplane $H_{yz}=\{x \in \re^k \mid (y-x)^2=(z-x)^2\}$ such that each of the $(2n-1)$ ideal points, other than $z,$ belongs to the open half space of $H_{yz}$ which contains $y.$\\
Then, by Fact 1 above, $y \succ z.$\\
Case 2: Suppose $z \notin C.$ Then choose $w$ (in the boundary of) $C$ such that
$$w \in argmin_{x \in C} d(x,z).$$
Since the function $d(.,z)$ is continuous and $C$ is a compact subset of $Z,$ the existence of such a $w$ is ensured.\\
Consider the hyperplane $H_{wz}=\{x \in \re^k \mid (w-x)^2=(z-x)^2\}.$ Then by Fact 1, $w \succ z,$ as for each $i \in N,$ $\bar x_i$ lies in the open half-space of $H_{wz}$ that contains $w.$\\
If $w$ is some voter's ideal point (say, $\bar x_i$ for some $i \in N$) then, by the logic of Case 1 above, there exists some $y \in (int(C) \cap L_i)$ for which $y \succ w.$\\
Case 3: Suppose $z \in (C \setminus (\cup_{(i,i') \in \mu}L_i)).$ Then there exists at least one pair $(i, i') \in \mu$  such that $y,$ the point at which the perpendicular from $z$ on the straight line $L_i$ intersects $L_i,$ belongs to $int(C).$ Note that by Lemma 2, $y \succ z.$\\
Note that for each of the three Cases above, the desired $y$ is, in fact, in $int(C) \cap L_i$ for some $(i,i') \in \mu.$\\
Case 4: Otherwise, set $y=z.$ Notice that with this Case, $y={\bar x_i}$ for some $i \in N$ such that $y$ is not any extreme point of $C$ or $y \in L_i \cap (int(C))$ for some $(i,i') \in \mu.$\\

\noi The next Step of the proof starts with working with this point $y$ obtained in this Step.\\
If $x_0 \succ y$ then we are done. Otherwise we proceed to Step 2 below.\\

\noi {\em Step 2:} Recall that Step 1 above ensures the existence of a point $y$ such that for at least one pair $(i,i') \in mu,$ $y \in (C \cap L_i)$ (and such that either a finite sequence of majority-dominations reaches $y$ from $z$ or $y=z$).\\
Now consider the following Cases.\\
Case 1: Suppose that for some other pair $(j,j') \in (\mu \setminus \{i,i'\}),$ the angle between $L_i$ and $L_j$ is $\pi/2.$ Pick one such pair $(j,j').$ Then the perpendicular from $y$ on $L_j$  intersects $L_j$ precisely at $x_0.$ Then, by Lemma 2, $x_0 \succ y$ and the Proposition is proved.\\
Case 2: Suppose that for every pair $(j,j') \in (\mu \setminus \{i,i'\}),$ the angle between $L_i$ and $L_j,$ (say, $\angle {\bar x_i} x_0 {\bar x_j}$, by any necessary relabelling of the names of $i,i',j',j'$) is less than $\pi/2.$ For any point $w \in (L_i \cap C),$ for any $(j,j') \in \mu,$ by $\ell_{ij}(w)$ denote the line segment from $w$ perpendicular to $L_j.$ Similarly, for any point $q \in \ell_{ij}(w),$ (where $w \in (L_i \cap C)$ as in the previous sentence), by $\ell_{ji}(q)$ denote the line segment from $q$ perpendicular to $L_i.$\\
Since $y \in (L_i \cap C)$ is not an extreme point of $C,$ for at least one other different $(j,j') \in \mu,$ it must be true that $(\ell_{ij}(y)) \cap (int(C)) \neq \emptyset.$  Pick one such pair $(j,j').$\\
Now consider the following sequence of points:\\
\noi $w_0=y;$\\
\noi $w_1$ is the point at which the line segment $\ell_{ij}(w_0)$ intersects $L_j$ if this point of intersection is in $C;$ otherwise, $w_1$ is the point at which $\ell_{ij}(w_0)$ intersects the boundary of $C;$\\
\noi $w_2$ is the point at which the line segment $\ell_{ji}(w_1)$ intersects $L_i$ (note that $w_2$ must belong to $C$ as the line segment ${\overline{x_0w_0}} \in C$ and $\angle {\bar x_i} x_0 {\bar x_j}$ is less than $\pi/2$);\\
and likewise, for any odd positive integer $k,$
\noi $w_k$ is the point at which the line segment $\ell_{ij}(w_{k-1})$ intersects $L_j$ if this point of intersection is in $C;$ otherwise, $w_k$ is the point at which $\ell_{ij}(w_{k-1})$ intersects the boundary of $C;$\\
and for any even positive integer $k,$
\noi $w_k$ is the point at which the line segment $\ell_{ji}(w_{k-1})$ intersects $L_i.$\\
Note that by Lemma 2, for each positive integer $k,$ $w_k \succ w_{k-1}.$\\
Since the angle between $L_i$ and $L_j$ is less than $\pi/2,$ recall that for any even positive integer $k,$ $d(x_0, w_{k+2})<d(x_0, w_k).$\\ 
Now, for any $x \in Z,$ denote by $\rho_k(x)$ the point $y$ which is the reflection of $x$ by the straight line $L_k$ where $k \in \{i,j\}.$ Given the construction above, there exists an odd positive integer $m$ for which the following hold: for every odd $k \geq m,$\\
$(i)$ $w_k \in \overline{jj'};$ and\\
$(ii)$ each of the points $w_k,$ $\rho_i(w_k)$ as well as $\rho_j(\rho_i(w_k))$ is inside $B(x_0) \seq C$ where $B(x_0)$ is a ball centred at $x_0.$\\
If $x_0 \succ w_m$ then we are done. Otherwise we proceed to Step 3 below.\\ 

\noi {\em Step 3:} As in the previous Step, for any $x \in Z,$ denote by $\rho_k(x)$ the point $y \in {\re}^k$ which is the reflection of $x$ by the straight line $L_k$ where $k \in \{i,j\}.$ Consider the following sequence of points:{\footnote{Note that if $Z={\re}^k,$ instead of being a proper subset of it, then Step 2 of the proof can be skipped and Step 3 can start by setting $y_0=y.$}\\
\noi $y_0=w_m,$ (as defined in the Step 2 above);\\
\noi $y_1=\rho_i(y_0);$\\
\noi $y_2=\rho_j(y_1);$\\
and likewise, for any odd positive integer $k,$
\noi $y_k=\rho_i(y_{k-1});$\\
and likewise, for any even positive integer $k,$
\noi $y_k=\rho_j(y_{k-1}).$\\
Note that each point in the sequence $\{y_k\}$ is obtained by rotation centred at $x_0.$ Also note that for each $k,$ for any $x$ belonging to the open line segment ${\overline{y_ky_{k+1}}} \setminus \{y_k,y_{k+1}\},$ by Lemma 2, $x \succ y_k.$\\
Suppose, without loss of generality, $y_0 \in C_S(x_0)$ (recall that if $y_0 \in (Z \setminus M(x_0)$ then we are already done) for some $n$-voter coalition $S.$ Then construct, mimicking the construction above, a slightly modified sequence of points is constructed as follows:\\   
\noi $q_0=w_m,$ (as defined in the Step 2 above);\\
\noi $q_1$ is a point on the line-segment ${\overline{q_0 \rho_i(q_0)}}$ infinitesimally close to $\rho_i(q_0);$\\
\noi $q_2$ is a point on the line-segment ${\overline{q_1 \rho_j(q_1)}}$ infinitesimally close to $\rho_j(q_1);$\\
and likewise, for any odd positive integer $k,$
\noi $q_k$ is a point on the line-segment ${\overline{q_{k-1} \rho_i(q_{k-1}}})$ infinitesimally close to $\rho_i(q_{k-1});$\\
and likewise, for any even positive integer $k,$
\noi $q_k$ is a point on the line-segment ${\overline{q_{k-1} \rho_i(q_{k-1}}})$ infinitesimally close to $\rho_j(q_{k-1}).$\\
Note that by Lemma 2, for each $k,$ $q_k \succ q_{k-1}.$ Note also that the points in this sequence $\{q_k\}$ forms a spiral of infinitesimal width centred at $x_0.$ If for any $k,$ $q_k \in Z \setminus M(x_0)$ then we are done. Suppose otherwise. Since $M(x_0)$ is star-shaped around $x_0,$ then it must be the case that for some $k,$ $q_{k-1} \in C_S(x_0)$ but $q_k \in C_T(x_0)$ for some $n$-voter coalition $T$ different from $S.$ Then, by Remark 1 above, ${\overline{q_{k-1}q_k}} \cap (Z \setminus M(x_0)) \neq \emptyset.$ Pick a point $p \in {\overline{q_{k-1}q_k}} \cap (Z \setminus M(x_0)).$ Then, by Lemma 2, $p \succ q_{k-1}$ and $x_0 \succ p.$  Then we are done.\qed\\

\section{A couple of concluding remarks}

Please note that if the policy space is one-dimensional then the core is singleton only if the single point in the core coincides with at least one of the ideal points. And of course, obtaining similar results with weaker assumptions is worth-exploring.

\section*{References}

\begin{description}

\item Austen-Smith, D., Banks, J., (1999). Positive Political Theory I, University of Michigan Press.

\item Bhattacharya, A., Ciardiello, F. (2024). How sensitive are the results in voting theory when just one other voter joins in? Some instances with spatial majority voting” (with Francesco Ciardiello). Discussion Papers in Economics, No. 24/03, University of York.

\item Cox, G. W., (1987). The uncovered set and the Core. American Journal of Political Science, 31: 408-422.

\item Davis, O. A., Degroot, M. H., Hinich, M. J. (1972). Social Preference Orderings and
Majority Rule. Econometrica, 40: 147-157.

\item Duggan, J., (2018). Necessary gradient restrictions at the core of a voting rule. Journal of Mathematical Economics, 79: 1-9.

\item Le Breton, M., Sen, A., 1999. Separable preferences, strategyproofness, and decomposability.  Econometrica, 67, 605-628.

\item McKelvey, R. D. (1976). Intransitivities in Multidimensional Voting Models and Some
Implications for Agenda Control. Journal of Economic Theory, 12: 472-482.

\item Miller, N. R., B. Grofman, S. L. Feld (1989). The Geometry of Majority Rule. Journal of Theoretical Politics, 1: 379-406.

\item Sengupta, A., Sengupta, K. (1996). A property of the core. Games and Economic Behavior, 12, 266-273.

\end{description}

\end{document}